\documentclass[aps, prd, floatfix, nofootinbib, superscriptaddress, twocolumn]{revtex4-2}

\usepackage{latexsym}
\usepackage{amsmath}
\usepackage{amssymb}
\usepackage{amsfonts}

\usepackage[mathscr,scaled=1.15]{urwchancal}
\DeclareFontFamily{OT1}{pzc}{}
\DeclareFontShape{OT1}{pzc}{m}{it}%
{<-> s * [1.15] pzcmi7t}{}
\DeclareMathAlphabet{\mathpzc}{OT1}{pzc}{m}{it}

\usepackage{CJKutf8}
\usepackage{color}
\usepackage{slashed}
\usepackage{dsfont}

\usepackage{supertabular}
\usepackage{placeins}
\usepackage{epsfig}
\usepackage{graphicx}

\definecolor{purple}{rgb}{0.5,0,0.5}
\definecolor{blue}{rgb}{0.0,0,0.9}
\definecolor{prdblue}{rgb}{0.133,0.118,0.498}
\usepackage[colorlinks=true, pdfstartview=FitV, linkcolor=prdblue, citecolor= prdblue, urlcolor=prdblue]{hyperref}

\begin{document}
\begin{CJK}{UTF8}{song}

\title{$\,$\\[-6ex]\hspace*{\fill}{\normalsize{\sf\emph{Preprint nos}.\
NJU-INP 082/23, USTC-ICTS/PCFT-23-40}}\\[1ex]
Nucleon to $\Delta$ axial and pseudoscalar transition form factors}

\author{Chen Chen
       $^{\href{https://orcid.org/0000-0003-3619-0670}{\textcolor[rgb]{0.00,1.00,0.00}{\sf ID}},}$}
\affiliation{Interdisciplinary Center for Theoretical Study, University of Science and Technology of China, Hefei, Anhui 230026, China}
\affiliation{Peng Huanwu Center for Fundamental Theory, Hefei, Anhui 230026, China}

\author{Christian S. Fischer%
       $^{\href{https://orcid.org/0000-0001-8780-7031}{\textcolor[rgb]{0.00,1.00,0.00}{\sf ID}},}$}
\affiliation{Institut f\"ur Theoretische Physik, Justus-Liebig-Universit\"at Gie{\ss}en, D-35392 Gie{\ss}en, Germany}
\affiliation{Helmholtz Forschungsakademie Hessen f\"ur FAIR (HFHF),
GSI Helmholtzzentrum f\"ur Schwerionenforschung, Campus Gie{\ss}en, 35392 Gie{\ss}en, Germany}

\author{Craig D. Roberts%
       $^{\href{https://orcid.org/0000-0002-2937-1361}{\textcolor[rgb]{0.00,1.00,0.00}{\sf ID}},}$}
\affiliation{School of Physics, Nanjing University, Nanjing, Jiangsu 210093, China}
\affiliation{Institute for Nonperturbative Physics, Nanjing University, Nanjing, Jiangsu 210093, China\\[0.5ex]
Email:
\href{mailto:chenchen1031@ustc.edu.cn}{chenchen1031@ustc.edu.cn} (CC);
\href{mailto:christian.fischer@theo.physik.uni-giessen.de}{christian.fischer@theo.physik.uni-giessen.de} (CSF);
\href{mailto:cdroberts@nju.edu.cn}{cdroberts@nju.edu.cn} (CDR)
}

\date{2023 December 21}

\begin{abstract}
A symmetry-preserving continuum approach to the calculation of baryon properties in relativistic quantum field theory is used to predict all form factors associated with nucleon--to--$\Delta$ axial and pseudoscalar transition currents, thereby unifying them with many additional properties of these and other baryons.  The new parameter-free predictions can serve as credible benchmarks for use in analysing existing and anticipated data from worldwide efforts focused on elucidation of $\nu$ properties.
\end{abstract}

\maketitle
\end{CJK}


\noindent\emph{1.\,Introduction}\,---\,%
With the discovery of neutrino ($\nu$) masses twenty-five years ago \cite{Super-Kamiokande:1998kpq, RevModPhys.88.030501, RevModPhys.88.030502}, science acquired tangible evidence of phenomena that lie beyond the Standard Model of particle physics (SM).  With such motivation, precision measurements of $\nu$, $\bar\nu$ (antineutrino) properties are today amongst the top priorities in fundamental physics \cite{Alvarez-Ruso:2017oui, Arguelles:2019xgp, DUNE:2020jqi, Formaggio:2021nfz, JUNO:2021vlw, Lokhov:2022zfn, Ruso:2022qes, SajjadAthar:2022pjt}.

Known neutrinos are very weakly interacting; so, their detection requires excellent understanding of detector responses to $\nu$ passage.  The array of interactions that can occur depends strongly on $\nu$ energy.  At low energies, elastic scattering on entire nuclei dominates; whereas at high energies, individual neutrons and protons (nucleons) are resolved and deep inelastic $\nu$-parton scatterings take place.
(Partons are the degrees-of-freedom used to define quantum chromodynamics -- QCD, \emph{i.e}., the SM theory of strong interactions.)
Yet another class of reactions is important in few-GeV reactor-based $\nu$ experiments, \emph{viz}.\ inelastic processes on single or small clusters of nucleons, which produce a range of excited-nucleon final states.

To some extent, the required cross-sections can be determined in dedicated $\nu$ scattering experiments. However, reliable strong interaction theory predictions are equally important.  In this context, results for nucleon ($N$) electroweak elastic and transition form factors become critical to interpreting modern $\nu$ oscillation experiments \cite{Mosel:2016cwa, Hill:2017wgb, Gysbers:2019uyb, Lovato:2020kba}.
Recognising this, the past five years have seen a focus on calculating the form factors associated with elastic $\nu N$ scattering using both continuum and lattice Schwinger function methods \cite{Chen:2021guo, ChenChen:2022qpy, Alexandrou:2017hac, Jang:2019vkm, Bali:2019yiy}.

However, as noted above, inelastic processes are also crucial to understanding existing and anticipated experiments.
The most significant of these are excitations of the $\Delta$-resonance via, \emph{e.g}., $\nu_\ell N \to \ell \Delta$ and $\nu_\ell N \to \ell (\pi N \to \Delta)$, where $\ell$ is a light lepton and the second process involves the $\pi$-meson \cite{Sato:2003rq, Simons:2022ltq}.  Only one set of related calculations exists \cite{Alexandrou:2007eyf, Alexandrou:2010uk}.  They were obtained using 
lattice-QCD (lQCD) with quark current masses that produce a pion mass-squared $m_\pi^2 \gtrsim 5 (m_\pi^{\rm empirical})^2$, $m_\pi^{\rm empirical}  = 0.135\,$GeV.  Lacking further studies, challenges remain, like extending computations to $m_\pi^{\rm empirical}$ and significantly reducing all uncertainties.

Hitherto, there have been no more recent predictions.  We address that lack herein, employing a continuum approach to the baryon bound-state problem that has widely been used with success \cite{Eichmann:2016yit, Barabanov:2020jvn, Ding:2022ows, Carman:2023zke}.  Of particular relevance are its unifying predictions for $N$ and $\Delta$-baryon elastic weak-interaction form factors \cite{Chen:2021guo, ChenChen:2022qpy, Yin:2023kom}.  Where comparisons with experiment and results obtained using other robust tools are available, there is agreement.






\medskip

\noindent\emph{2.\,Axial and pseudoscalar currents}\,---\,%
Weak $N\to \Delta$ transitions are described by the matrix element
\begin{subequations}
\begin{align}
{\mathcal J}_{5\mu}^{j}(K,Q&) := \langle\Delta(P_f;s_f)|{\mathcal A}^{j}_{5\mu}(0)|N(P_i;s_i)\rangle\\
&=
\bar{u}_\alpha(P_f;s_f)\Gamma_{5\mu,\alpha}^j(K,Q)u(P_i;s_i)\,,
\label{jaxdq0}
\end{align}
\end{subequations}
where $P_{i,j}(s_{i,j})$ are the momenta (spins) of the incoming nucleon and outgoing $\Delta$, respectively, which are on-shell: $P_{i,f}^2=-m_{N,\Delta}^2$; and $K=(P_f+P_i)/2$, $Q=P_f-P_i$.
Isospin symmetry is assumed throughout \cite{IZ80};
and the on-shell baryons are described by Dirac and Rarita-Schwinger spinors, respectively -- see Ref.\,\cite[Appendix\,B]{Segovia:2014aza} for details.
Introducing the column vector $\psi={\rm column}[u,d]$, with $u$, $d$ being the two light-flavour quark fields, then the axial current operator may be written
${\mathcal A}^{j}_{5\mu}(x)=\bar{\psi}(x)\frac{\tau^j}{2}\gamma_5\gamma_\mu\psi(x)$, in which the isospin structure is expressed via the Pauli matrices $\{\tau^j|j=1,2,3\}$.

Owing to the assumed isospin symmetry, a complete characterisation of $N\to\Delta$ weak transitions can be achieved by focusing solely on the neutral current ($j=3$) $p \to \Delta^+$ transition, in which case the vertex is typically written in the following form \cite{Adler:1968tw, LlewellynSmith:1971uhs}:
{\allowdisplaybreaks
\begin{align}
\nonumber
\Gamma_{5\mu,\alpha}^{3}(K,Q) &= \sqrt{\tfrac{2}{3}}\bigg[i(\gamma_\mu Q_\alpha-\delta_{\mu\alpha}\slashed{Q}) C^A_3(Q^2)/m_N\\
\nonumber
&-(\delta_{\mu\alpha}(P_f\cdot Q)-{P_f}_\mu Q_\alpha) C^A_4(Q^2)/m_N^2\\
&+\delta_{\mu\alpha}C^A_5(Q^2) - Q_\mu Q_\alpha  C^A_6(Q^2)/m_N^2\bigg]\,,
\label{AxialCurrent}
\end{align}
where $C^A_{3,4,5,6}$ are the four Poincar\'e-invariant transition form factors and $\sqrt{2/3}$ is the isospin coefficient. 
}

The related pseudoscalar transition matrix element is
\begin{subequations}
\begin{align}
{\mathcal J}_{5}^j(K,Q&) = \langle\Delta(P_f;s_f)|{\mathcal	 P}^j_{5}(0)|N(P_i;s_i)\rangle\\
&=\bar{u}_\alpha(P_f;s_f)\Gamma_{5,\alpha}^j(Q)u(P_i;s_i)\,,
\end{align}
\end{subequations}
where ${\mathcal P}^{j}_{5}(x)=\bar{\psi}(x)\frac{\tau^j}{2}\gamma_5\psi(x)$.
For the neutral current:
\begin{align}
\Gamma_{5,\alpha}^{3}(Q) &= \sqrt{\tfrac{2}{3}}
i\frac{Q_\alpha}{4m_N}\frac{m_\pi^2}{Q^2+m_\pi^2}\frac{f_\pi}{m_q}G_{\pi N\Delta}(Q^2) \,. \label{jpsdq0}
\end{align}
Here, $G_{\pi N\Delta}$ is the pseudoscalar transition form factor, $m_q$ is the light-quark current-mass, and $f_\pi=0.092\,$GeV is the pion leptonic decay constant \cite{Workman:2022ynf}.

Using the PCAC (partially conserved axial current) identity \cite{IZ80}:
$\partial_\mu{\mathcal A}^j_{5\mu}(x) + 2m_q{\mathcal P}^j_{5}(x)=0$,
one obtains the following ``off-diagonal'' PCAC relation:
\begin{equation}
\label{pcacff}
C^A_5(Q^2) - \frac{Q^2}{m_N^2}C^A_6(Q^2) = \frac{f_\pi m_\pi^2/[2 m_N]}{Q^2+m_\pi^2}G_{\pi N\Delta}(Q^2)\,.
\end{equation}
Drawing a parallel with the nucleon \cite{Chen:2021guo}, $C^A_5$ is kindred to the nucleon's axial form factor, $G_A$, and $C^A_6$ to its induced-pseudoscalar form factor, $G_P$.
Since $\left.Q^2C^A_6(Q^2)\right|_{Q^2=0}=0$, one immediately obtains the off-diagonal Goldberger-Treiman (GT) relation
\begin{align}
\label{gtr}
2 m_N C^A_5(0) = f_\pi G_{\pi N\Delta}(0)\,.
\end{align}
As consequences of chiral symmetry and its breaking pattern, only frameworks preserving Eqs.\,\eqref{pcacff}, \eqref{gtr} are viable.



\medskip

\begin{figure}[t]
\centerline{%
\includegraphics[clip, width=0.425\textwidth]{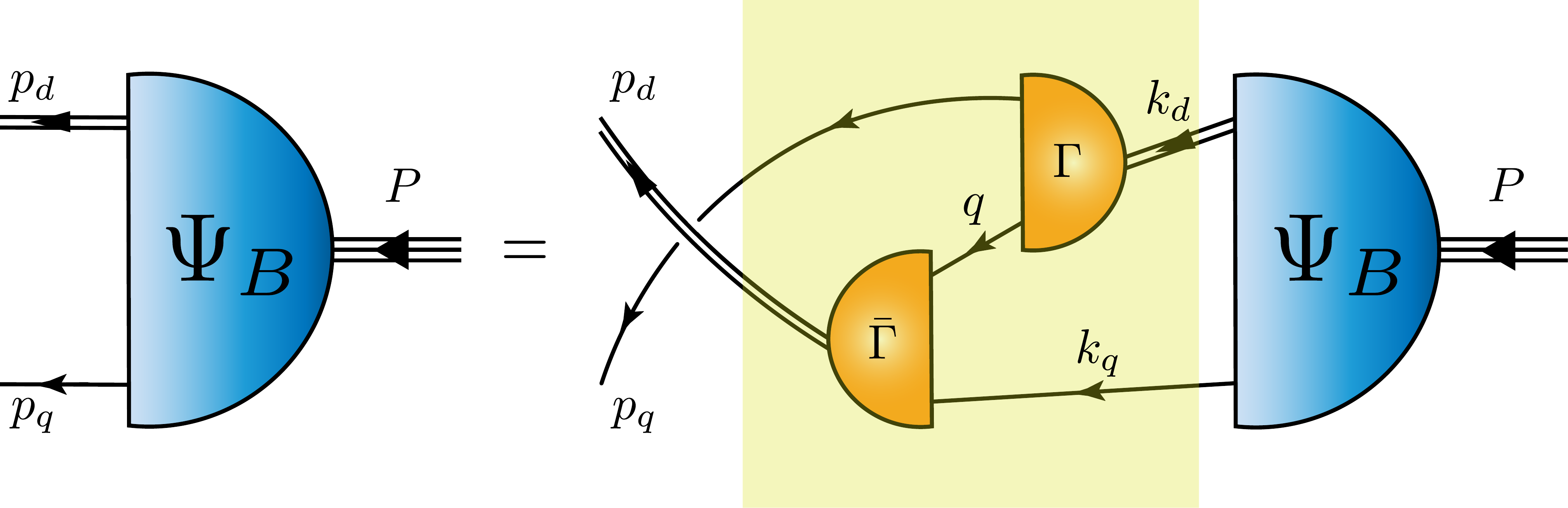}}
\caption{\label{figFaddeev}
Faddeev integral equation, whose solution is the Poincar\'e-covariant matrix-valued function $\Psi_B$, the Faddeev amplitude for a baryon $B$ with total momentum $P=p_q+p_d$.
Legend. \emph{Shaded box } -- Faddeev kernel; \emph{single line} -- dressed-quark propagator; $\Gamma$ -- diquark correlation amplitude; and \emph{double line} -- diquark propagator.
Only isoscalar-scalar diquarks, $[ud]$, and isovector--axialvector diquarks, $\{uu\}$, $\{ud\}$, $\{dd\}$ play a material role in nucleons and $\Delta$-baryons \cite{Barabanov:2020jvn}.}
\end{figure}

\noindent\emph{3.\,Baryon structure}\,---\,%
%
The structure of any given baryon can be described by a Faddeev amplitude, obtained from a Poincar\'e-covariant Faddeev equation that sums all possible exchanges and interactions which can occur between the three dressed-quarks that characterise its valence content.
Studies of the associated scattering problem reveal that there is no interaction contribution from the diagram in which each leg of the three-gluon vertex ($3g$V) is attached to a separate quark \cite[Eq.\,(2.2.10)]{Barabanov:2020jvn}.
Thus, whilst the $3g$V is a primary factor in generating the process-independent QCD effective charge \cite{Cui:2019dwv, Deur:2023dzc}, its role within baryons is largely to produce tight quark + quark (diquark) correlations.
Consequently, the baryon bound-state problem may reliably be transformed \cite{Cahill:1988dx, Reinhardt:1989rw, Efimov:1990uz} into solving the homogeneous matrix equation in Fig.\,\ref{figFaddeev}.

Each line and vertex in Fig.\,\ref{figFaddeev} is specified in Ref.\,\cite{Chen:2021guo}, which delivers parameter-free predictions for the nucleon $G_{A,P}$ form factors and unifies them with the pion-nucleon form factor, $G_{\pi NN}$.  A key to the success of that study is a sound expression of emergent hadron mass and its corollaries \cite{Roberts:2021nhw, Binosi:2022djx, Salme:2022eoy, Ding:2022ows, deTeramond:2022zcm, Ferreira:2023fva, Krein:2023azg}, such as a running light-quark mass whose value at infrared momenta, $M_D \approx 0.4\,$GeV, defines the natural magnitude for mass-dimensioned quantities in the light-quark sector of the Standard Model.  Associated mass-scales for the isoscalar-scalar and isovector-axialvector diquarks are [in GeV]:
\begin{align}
\label{dqmass}
m_{[ud]} = 0.80\,,\,\,\,
m_{\{uu\}} = m_{\{ud\}} = m_{\{dd\}} = 0.89\,.
\end{align}

In computing all transition form factors, we follow Refs.\,\cite{Chen:2021guo, ChenChen:2022qpy, Yin:2023kom}:  they are obtained from the nucleon elastic weak form factor expressions by replacing all inputs connected with the final state by those for the $\Delta$-baryon.  The normalisations of the $N$ and $\Delta$ Faddeev amplitudes are available from Refs.\,\cite{Chen:2021guo, Yin:2023kom}.  (The interaction current is included with the supplemental material.)

%




\medskip

\begin{figure}[t]
\vspace*{1.2em}

\leftline{\hspace*{0.5em}{\large{\textsf{A}}}}
\vspace*{-5ex}
\includegraphics[clip, width=0.43\textwidth]{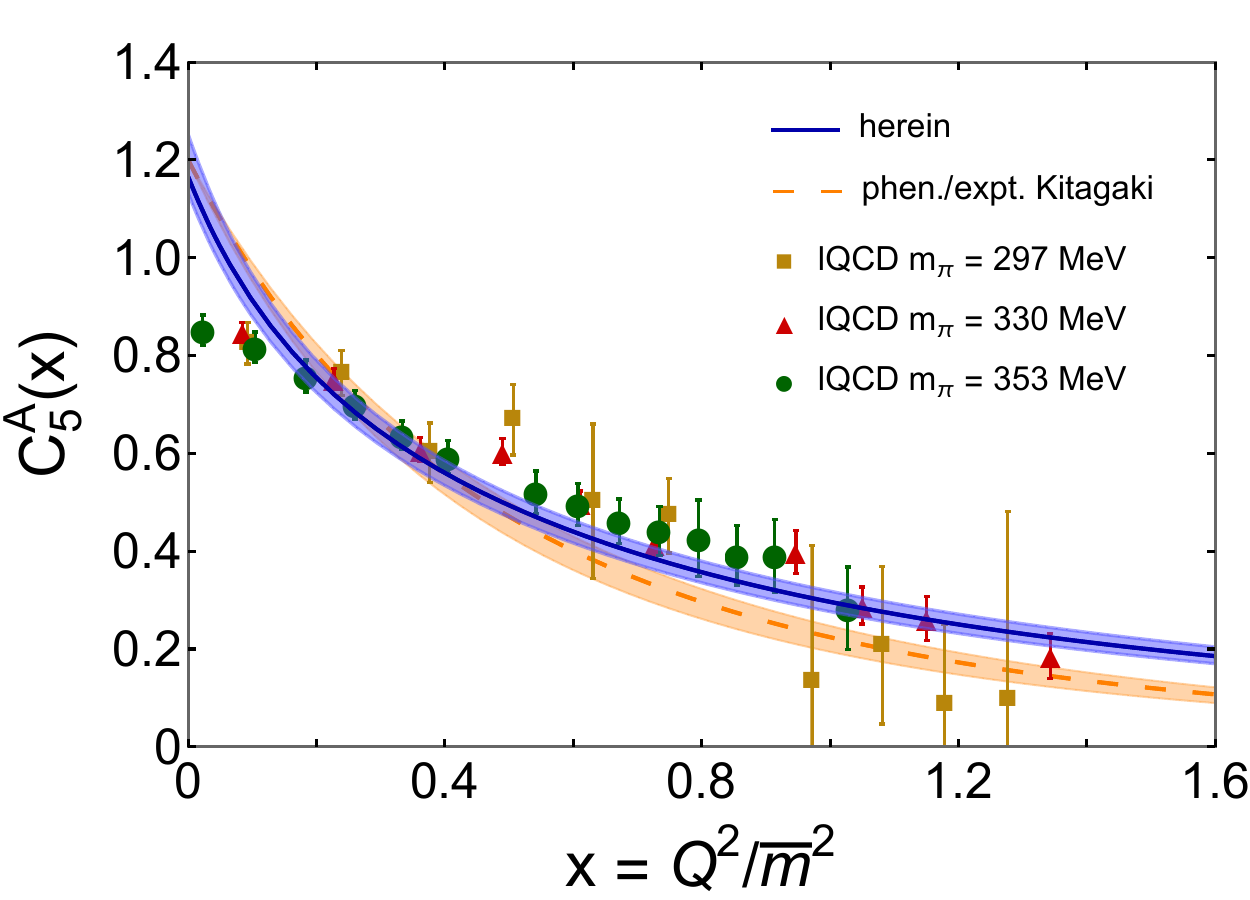}
\vspace*{1ex}
\leftline{\hspace*{0.5em}{\large{\textsf{B}}}}
\vspace*{-5ex}
\includegraphics[clip, width=0.43\textwidth]{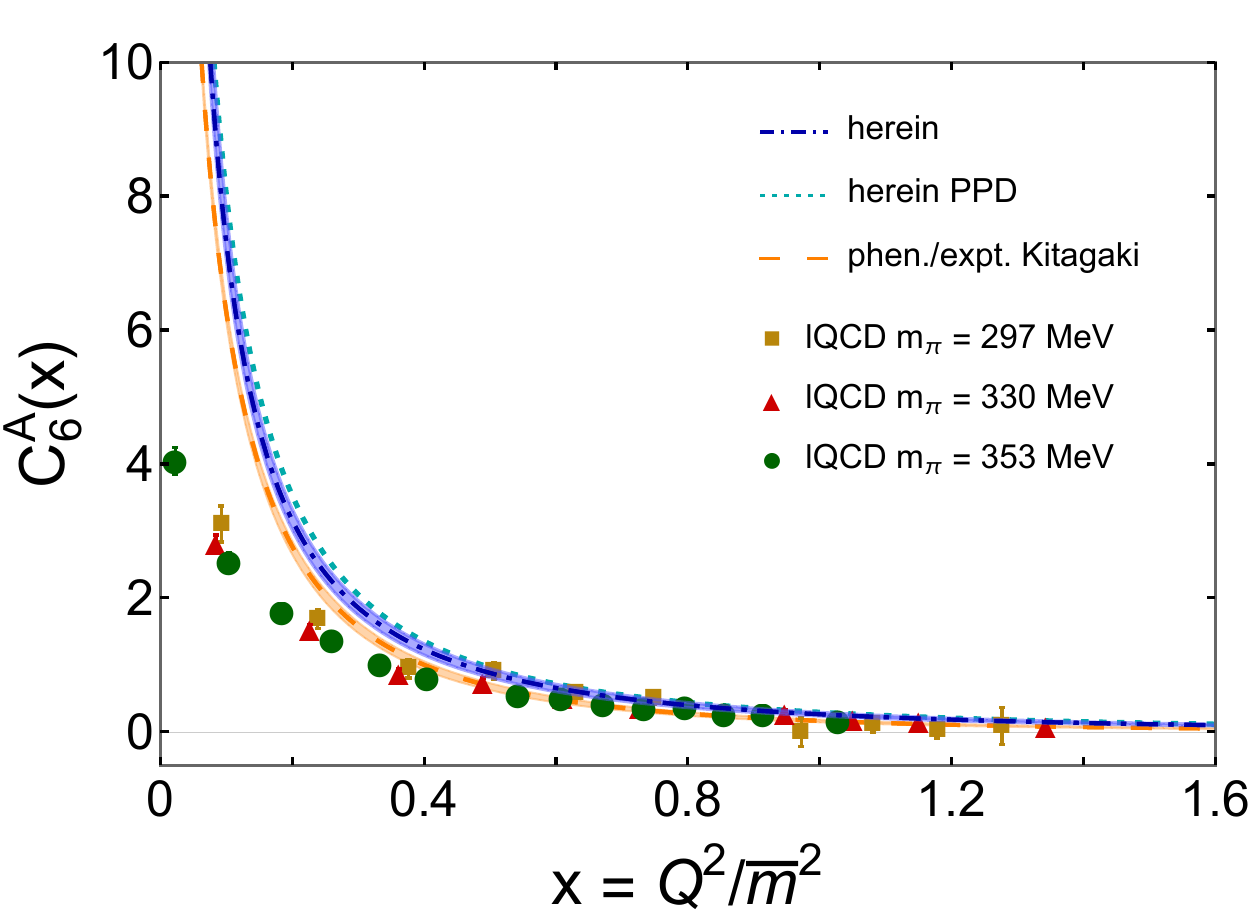}
\vspace*{4.5ex}
\caption{\label{FigCA56x}
{\sf Panel A}.  $C^A_5$ calculated herein -- blue curve within lighter blue uncertainty band; and empirical result \cite[Kitagaki]{Kitagaki:1990vs} -- dashed orange curve within uncertainty band.
{\sf Panel B}.
$C^A_6$ calculated herein -- dot-dashed blue curve within uncertainty band;
dotted cyan curve -- PPD approximation, Eq.\,\eqref{ppd}, using our prediction for $C^A_5(x)$;
and dashed orange curve within uncertainty band -- PPD approximation obtained with empirical result \cite{Kitagaki:1990vs} for $C^A_5(x)$.
Both panels:
lQCD \cite{Alexandrou:2007eyf, Alexandrou:2010uk} --
hybrid (HYB) [green circles -- $m_\pi=353$\,MeV] and
domain wall fermions (DWF) [gold squares -- $m_\pi=297$\,MeV, red triangles -- $m_\pi=330$\,MeV].
(Parametrisations of our form factor predictions are provided in the supplemental material.)
}
\end{figure}

\noindent\emph{4.\,Dominant $N\to \Delta$ transition axial form factors}\,---\,%
%
%
$C^A_{5,6}$ in Eq.\,\eqref{AxialCurrent} 
are dominant in the axial transition current.
Our prediction for $C^A_{5}(x=Q^2/\bar m^2)$, $\bar m = [m_N + m_\Delta]/2$, is drawn in Fig.\,\ref{FigCA56x}A.  On the domain depicted, $C^A_5(x)$ can fairly be interpolated using a dipole characterised by the $x=0$ value and axial mass listed in the first row of Table~\ref{tabQ20}.
The $x=0$ value is $0.94(5)$-times the prediction for $g_A$, the nucleon axial coupling, as calculated using the same framework;
and the axial mass is $0.95(3)$-times the analogous scale in a dipole representation of $G_A(x)$, the nucleon axial form factor, \emph{viz}.\ the axial transition form factor is softer, something also found in a coupled channels analysis of axial $N\to \Delta$ transitions \cite{Sato:2003rq}.

The reported uncertainty in our predictions expresses the impact of $\pm 5\%$ variations of the diquark masses in Eq.\,\eqref{dqmass}.  The results obtained from independent variations are combined with weight fixed by the relative strength of scalar $(0^+)$ and axialvector $(1^+)$ diquark contributions to $C^A_5(0)$, \emph{i.e.}, approximately $4$:$1$.
Like the nucleon case \cite{Chen:2021guo, ChenChen:2022qpy}, the $0^+$ and $1^+$ mass variations interfere destructively, \emph{e.g.}, reducing $m_{[ud]}$ increases $C^A_5(0)$, whereas $C^A_5(0)$ decreases with the same change in $m_{1^+}$.

In phenomenology, following Ref.\,\cite{Schreiner:1973mj}, it has been common to represent $C^A_{5}$ as follows \cite{Kitagaki:1990vs}:
\begin{align}
\label{adlerffs}
C^A_5(Q^2) = C^A_5(0)\frac{[1+ a_5 Q^2/(b_5 + Q^2)]}{(1+Q^2/\tilde m_A^2)^{2}}\,,
\end{align}
where $C^A_5(0)=1.2$, $a_5=-1.21$, $b_5=2\,$GeV$^2$ are parameters discussed in Ref.\,\cite{Schreiner:1973mj}.  The value of $C^A_5(0)$ is an estimate based on the off-diagonal GT relation -- Eq.\,\eqref{gtr}.   Using Eq.\,\eqref{adlerffs} and the same functional form for $C^A_{3,4,6}$, then a fit to $\nu d \to \mu^- \Delta^{++} n$ bubble chamber data yielded the value of $\tilde m_A$ in Table~\ref{tabQ20}.

\begin{table}[t]
\caption{\label{tabQ20}
Selected characteristics of $C^A_5$ compared with combined phenomenology and experiment \cite{Kitagaki:1990vs} and other available calculations \cite{Alexandrou:2007eyf, Alexandrou:2010uk}.
($2 \bar m = m_N + m_\Delta$.)
}
\begin{tabular*}
{\hsize}
{
l@{\extracolsep{0ptplus1fil}}
l@{\extracolsep{0ptplus1fil}}
l@{\extracolsep{0ptplus1fil}}}\hline\hline
 & $C^A_5(0)$ & $\tilde m_A/\bar{m}$ \\\hline
%
Herein & $1.16^{+0.09}_{-0.03}$ & $1.01(2)$ \\\hline
%
Phenomenology/Empirical \cite{Kitagaki:1990vs} & $1.2$ & $1.18(8)$ \\
Phenomenology/Empirical (dipole)  & $1.2$ & $0.88(4)$ \\\hline
%
%
lQCD\,\cite{Alexandrou:2007eyf} $m_\pi =353$\,MeV & $0.90(1)$ & $1.32(3)$ \\
lQCD\,\cite{Alexandrou:2010uk} $m_\pi = 297$\,MeV & $0.94(6)$ & $1.32(13)$ \\
lQCD\,\cite{Alexandrou:2010uk} $m_\pi=330$\,MeV & $0.97(3)$ & $1.23(5)$ \\
\hline\hline
\end{tabular*}
\end{table}

Taken at face value, this result is $1.34(17)$-times that which some phenomenological estimates now associate with the nucleon axial form factor \cite{Meyer:2016oeg}.  However, as noted above, such estimates are based on dipole parametrisations.  If one recasts Eq.\,\eqref{adlerffs} as a dipole, then the empirical fit corresponds to $\tilde m_A = 1.01(5)m_N=0.878(41)\bar m$ -- we have listed this value in Row~3 of Table~\ref{tabQ20}.  Using an ${\mathpzc L}_1$ measure, then, on the phenomenological fitting domain $Q^2\in [0,3]\,$GeV$^2$, the curves thus obtained agree with the original parametrisations to within 7.6(1.1)\%.  This dipole mass scale is $0.70(3)$-times that associated with the nucleon $G_A$.
%

Our prediction for $C^A_5$ in Fig.\,\ref{FigCA56x}A is contrasted therein with a phenomenological/empirical result \cite{Kitagaki:1990vs} and lQCD computations \cite{Alexandrou:2007eyf, Alexandrou:2010uk}.  Allowing for the constraint imposed on data analysis by working with Eq.\,\eqref{adlerffs}, the existing phenomenological extraction agrees well with our prediction.
Regarding the lQCD results, however, there are significant differences.  Compared to our prediction and phenomenology, the lQCD results lie lower on $x\simeq 0$ and are harder -- these differences are quantified in Table~\ref{tabQ20}.  It seems fair to judge that these discrepancies owe primarily to the larger than physical pion masses which characterise the lattice configurations.


We draw our prediction for $C^A_6(x)$ in Fig.\,\ref{FigCA56x}B, wherein it is compared with
available lQCD computations.
%
In this case, too, the lQCD results lie low on $x \lesssim 0.4$ and are harder than our prediction.

Since $C^A_6$ is kindred to the nucleon induced pseudoscalar form factor, $G_P$, it is natural to expect that a pion pole dominance (PPD) approximation is also valid for the $N$-$\Delta$ axial transition.  Referring to Eq.\,\eqref{pcacff}, this off-diagonal PPD approximation is ($\mu_N\equiv m_N^2/\bar{m}^2$, $\mu_\pi\equiv m_\pi^2/\bar{m}^2$)
\begin{align}
\label{ppd}
C^A_6(x) \approx C^A_5(x) \mu_N/[x+\mu_\pi]\,.
\end{align}
Figure~\ref{FigCA56x}B reveals it to be a quantitatively reliable association, just as its analogues are for the nucleon and $\Delta$-baryon weak elastic form factors \cite{Chen:2021guo, Yin:2023kom}.  (Additional information is provided in the supplemental material.)




\begin{figure}[t]
\vspace*{1.2em}

\leftline{\hspace*{0.5em}{\large{\textsf{A}}}}
\vspace*{-5ex}

\hspace*{-2ex}\includegraphics[clip, width=0.44\textwidth]{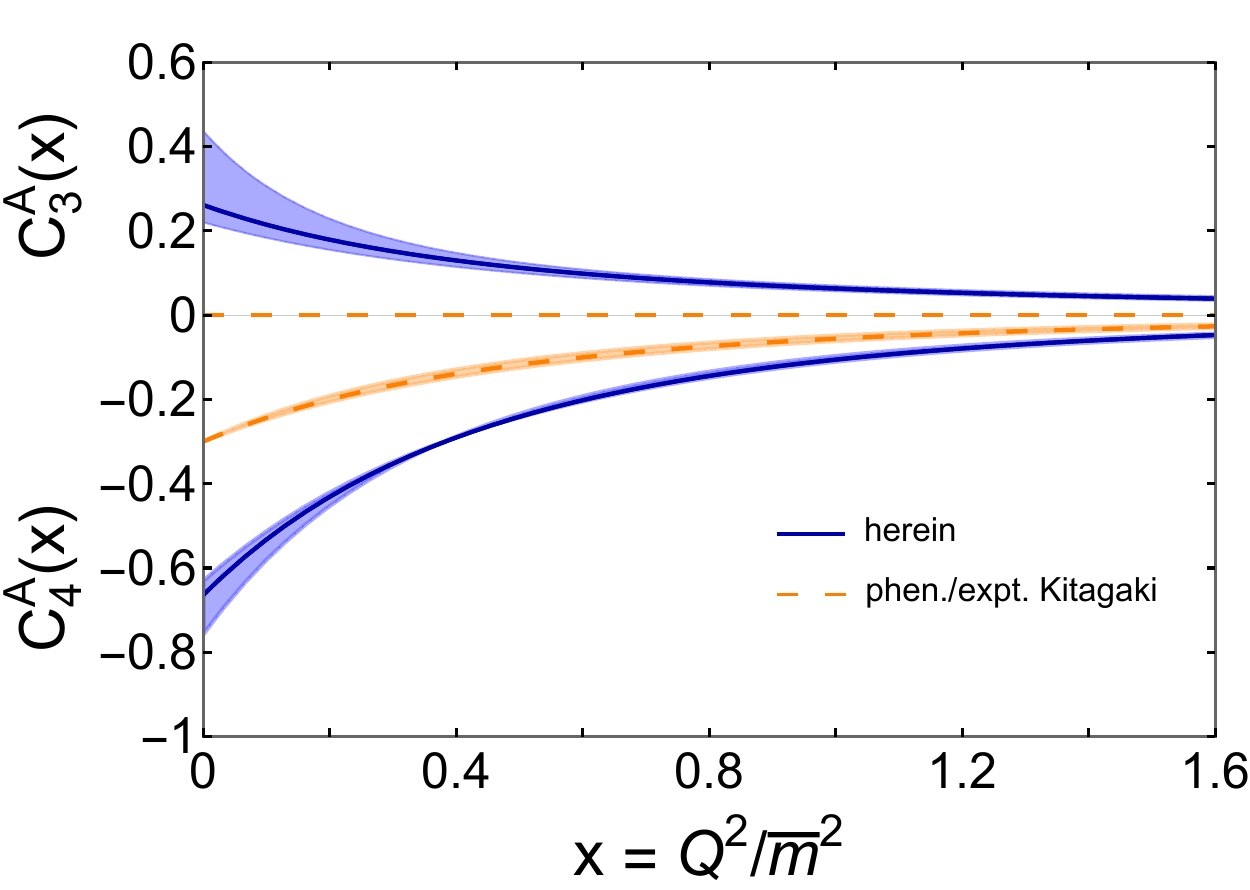}
\vspace*{0.5ex}
\leftline{\hspace*{0.5em}{\large{\textsf{B}}}}
\vspace*{-5ex}
\includegraphics[clip, width=0.425\textwidth]{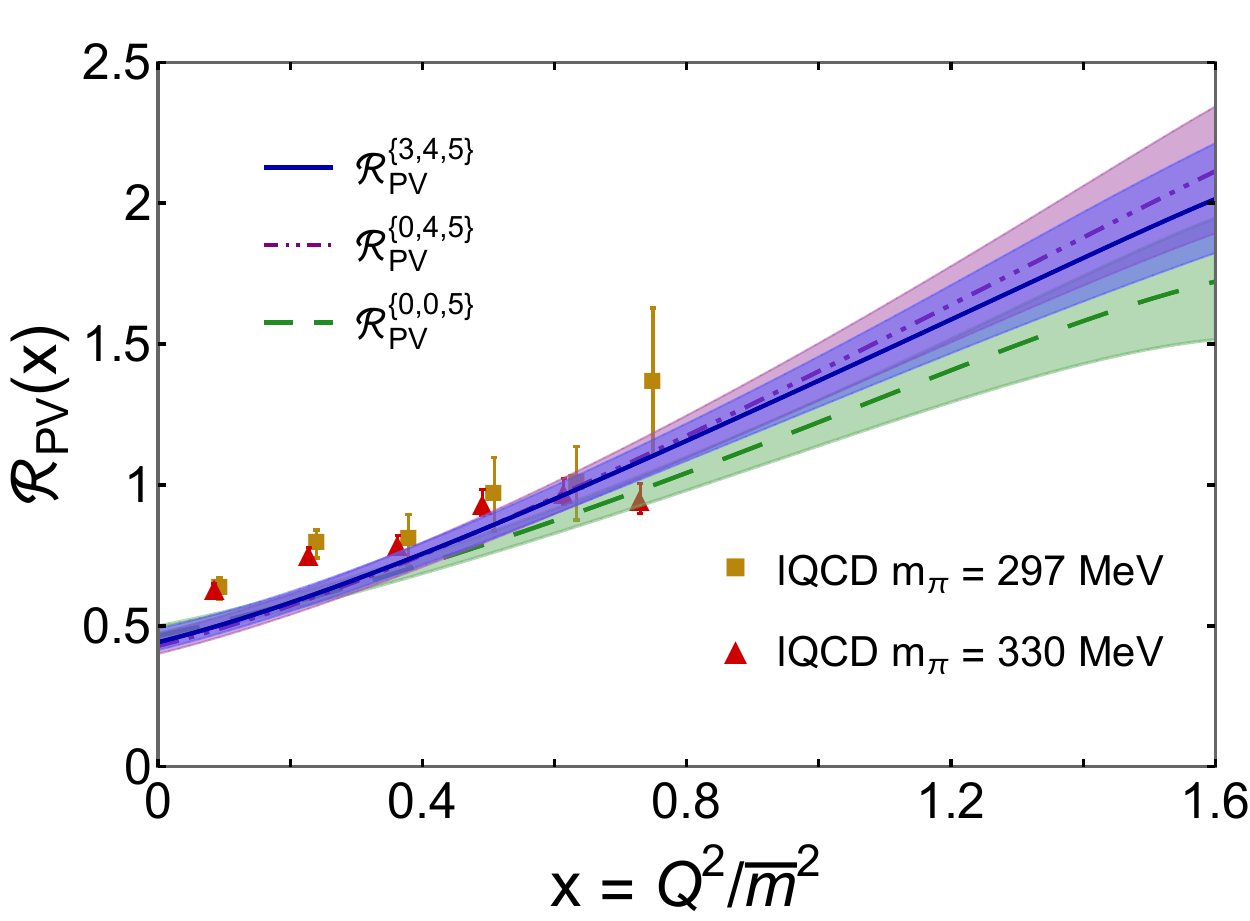}
\vspace*{4.5ex}
%
%
\caption{\label{FigCA34x}
{\sf Panel A}.
Predictions for subdominant transition form factors -- blue curves within bracketing uncertainty band:
$C^A_4(x)$, negative curves; and $C^A_3(x)$, non-negative curves.
Both cases: empirical extraction \cite[Kitagaki]{Kitagaki:1990vs} -- dashed orange curve and bracketing band.
%
{\sf Panel B}.
PV ratio, Eq.\,\eqref{rpv345}: legend superscript indicates transition form factors retained in the calculation.
For comparison, lQCD results \cite{Alexandrou:2010uk}: gold squares -- $m_\pi=297$\,MeV, red triangles -- $m_\pi=330$\,MeV.
}
\end{figure}

%

\medskip

\noindent\emph{5.\,Subdominant $N\to\Delta$ transition form factors}\,---\,
%
Our predictions for $C^A_{3,4}$ are drawn in Fig.\,\ref{FigCA34x}A and compared with the phenomenological extractions \cite{Kitagaki:1990vs}: the models used therein assume $C^A_3 \equiv 0$ and $C^A_4=-C^A_5/4$.  Our predictions disagree markedly with those assumptions.

$C^A_3$ is the weak $N\to \Delta$ analogue of the electric quadrupole ($E2$) form factor in $\gamma N \to \Delta$ transitions \cite{Liu:1995bu, Barquilla-Cano:2007vds}.
Like the vector $E2$ transition strength, if the nucleon and $\Delta$-baryon are described by SU$(6)$ symmetric (spherical) wave functions, then $C^A_3 \equiv 0$.  Evidently, in our Poincar\'e covariant treatment, $C^A_3 \neq 0$.  Indeed, no Poincar\'e-covariant $N$ or $\Delta$-baryon wave function is simply $\mathsf S$-wave in character -- see, \emph{e.g}., the wave functions in Ref.\,\cite{Liu:2022ndb}.
We find $C^A_3(x)$ decreases monotonically with increasing $x$ from a maximum value $C^A_3(0) = 0.26^{+0.17}_{-0.04}$.



Considering Fig.\,\ref{FigCA34x}A, our prediction for $C^A_4$ exhibits qualitatively similar behaviour to the phenomenological parametrisation. 
Quantitatively, however, it is uniformly larger in magnitude: we predict $C^A_4(0)=-0.66^{+0.03}_{-0.10}$, versus $C^A_4(0)=-0.3$ in the parametrisation.



\medskip

\noindent\emph{6.\,Parity violation asymmetry}\,---\,
With the construction and use of high-luminosity facilities, spin observables can today be used to probe hadron structure and search for beyond-SM physics -- see, \emph{e.g}., Refs.\,\cite{G0:2011rpu, Becker:2018ggl, JeffersonLabSoLID:2022iod}.
One such quantity is the parity-violating asymmetry in electroweak excitation of the $\Delta$-baryon \cite{Mukhopadhyay:1998mn}:
{\allowdisplaybreaks
\begin{align}
\nonumber
{\mathcal R}_{PV}^{\{3,4,5\}}(&Q^2):=\frac{C^A_5}{C^V_3}\bigg[1+\frac{m_\Delta^2-Q^2-m_N^2}{2m_N^2}\frac{C^A_4}{C^A_5}\\
&-\frac{m_N^2+Q^2+2m_Nm_\Delta-3m_\Delta^2}{4m_Nm_\Delta}\frac{C^A_3}{C^A_5}\bigg]\,.
\label{rpv345}
\end{align}
Having calculated the weak $N\to \Delta$ transition form factors, in Fig.\,\ref{FigCA34x}B we deliver a prediction for this as yet unmeasured ratio.
Since $\pi N$ final-state interactions play a material role in understanding the low-$x$ behaviour of $\gamma N \to \Delta$ transition form factors \cite{Julia-Diaz:2006ios} and such effects are difficult to represent reliably -- see, \emph{e.g}., Ref.\,\cite[Sec.\,6]{Segovia:2014aza}, we
use our predictions for $C_{3,4,5}^A$ but construct $C^V_3$ from the parametrisations 
presented in Ref.\,\cite[MAID]{Drechsel:2007if}.
}

Two principal conclusions may be drawn from Fig.\,\ref{FigCA34x}B:
(\emph{i}) ${\mathcal R}_{PV}(0) = 0.45(4) >0$, which may be compared with a value of $\approx 0.6$ if one uses the empirical parametrisations from Ref.\,\cite{Kitagaki:1990vs}; and
(\emph{ii}) $C^A_5$ is dominant on $x\lesssim 0.5$, but $C^A_{3,4}$ become increasingly important as $x$ grows.

\medskip

\begin{figure}[t]
\centerline{%
\includegraphics[clip, width=0.425\textwidth]{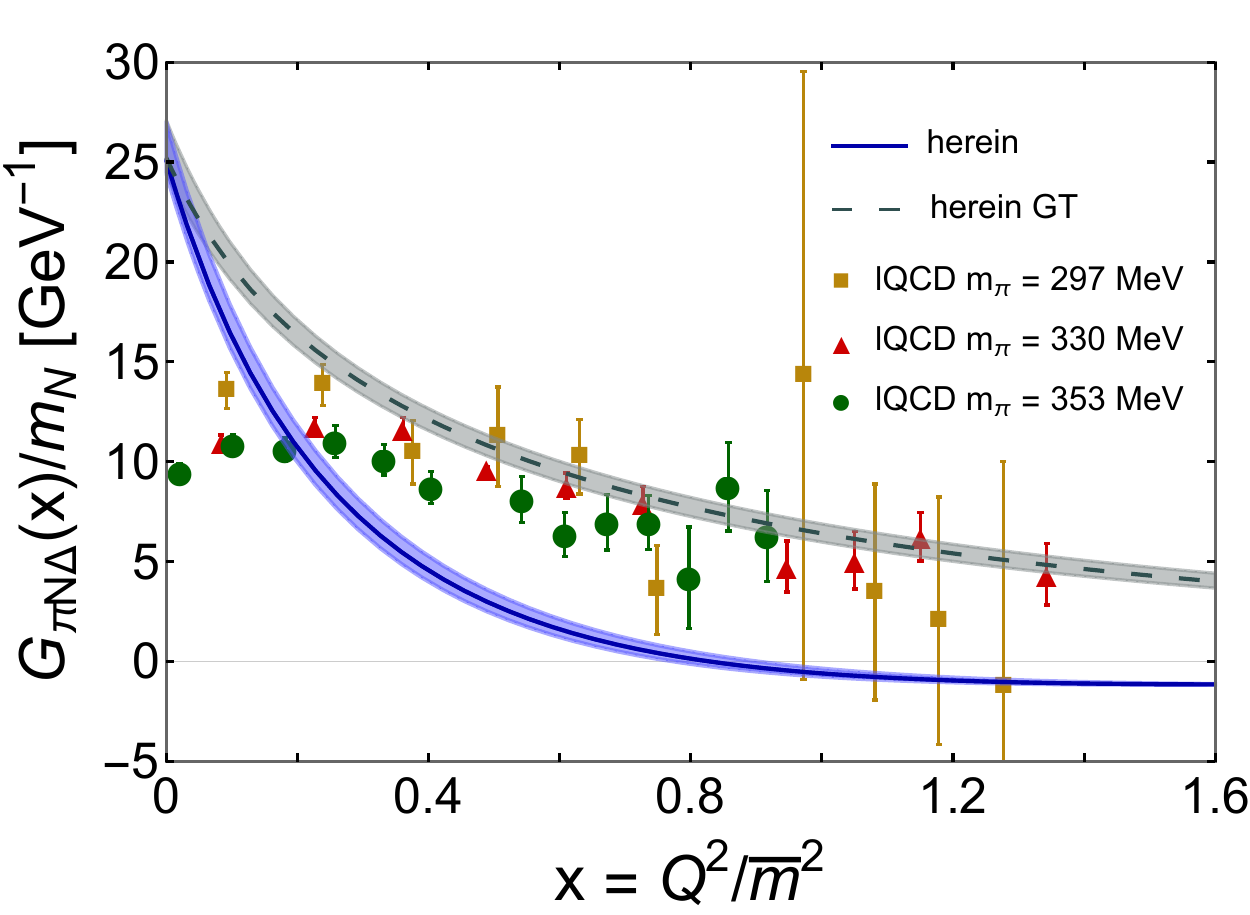}}
\caption{\label{figgpind}
$G_{\pi N\Delta}(x)/m_N$ calculated herein -- blue curve within uncertainty band;
$G'_{\pi N\Delta}(x)/m_N$ -- dashed grey curve and band, where our result for $C^A_5(x)$ is used on the right in Eq.\,\eqref{psgt}.
%
Comparison:
lQCD \cite{Alexandrou:2007eyf, Alexandrou:2010uk} --
hybrid [green circles -- $m_\pi=353$\,MeV] and
domain wall fermions [gold squares -- $m_\pi=297$\,MeV, red triangles -- $m_\pi=330$\,MeV].
Our prediction for $G_{\pi N\Delta}(-m_\pi^2)/m_N = 27.43^{+2.02}_{-0.68}$ is consistent with that obtained using chiral perturbation theory \cite{Unal:2021byi}.
}
\end{figure}

\noindent\emph{7.\,Pseudoscalar transition form factor}\,---\,
Our prediction for $G_{\pi N\Delta}$ is drawn in Fig.\,\ref{figgpind}.
Defined as usual, the $G_{\pi N\Delta}$ radius is
$r_{\pi N\Delta} =1.4(1)\,r_{\pi NN}$,
where $r_{\pi NN}$ is the $G_{\pi NN}$ analogue \cite{Chen:2021guo}.
Our prediction is softer than available lQCD results.
Notably, owing to a persistent negative contribution from probe seagull couplings to the diquark-quark vertices, we find a zero in $G_{\pi N\Delta}$ at $x =  0.84(6)$.  (See Fig.\,\ref{figcurrent} and Table~\ref{tabQ20sep} in the supplemental material.)  Given the large pion masses and poor precision, extant lQCD results cannot test this prediction.

It is clear from Fig.\,\ref{figgpind} that the GT relation, Eq.\,\eqref{gtr}, is satisfied: specifically [in $1/$GeV],
\begin{equation} 
\frac{G_{\pi N\Delta}(0)}{m_N} = 25.2^{+1.9}_{-0.7} \; {\rm cf.} \;
\frac{2C^A_5(0)}{f_\pi} = 25.1^{+1.9}_{-0.6} \,.
\label{GTtest}
\end{equation}
As with nucleon and $\Delta$ elastic weak form factors, the GT relation is only valid on $x\simeq 0$.
Symmetry guarantees nothing more.  We highlight this in Fig.\,\ref{figgpind} by also plotting
\begin{align}
\label{psgt}
G^\prime_{\pi N\Delta}(x)/m_N:= (2/f_\pi ) C^A_5(x)\,.
\end{align}
Plainly, $G^\prime_{\pi N\Delta}(x) \approx G_{\pi N\Delta}(x)$ only on $x\lesssim 0.1$.
(Additional remarks on PCAC are contained in the supplemental material.)




\medskip

\noindent\emph{8.\,Summary}\,---\,
Using a Poincar\'e-covariant, symmetry-preserving treatment of the continuum baryon bound-state problem, in which all elements possess an unambiguous link with analogous quantities in QCD, we delivered parameter-free predictions for every form factor associated with $N\to \Delta$ transitions driven by axial or pseudoscalar probes.  In doing so, we completed a unified description of, \emph{inter alia}, all $N$ and/or $\Delta$-baryon axialvector and pseudoscalar currents.  Where comparisons with data are available, our predictions are confirmed.  Hence, the results herein can serve as a reliable resource for use in analysing existing and anticipated data relevant to worldwide efforts focused on elucidation of $\nu$, $\bar \nu$ properties.
No material improvement upon our results may be envisaged before the same array of observables can be calculated using either the explicit three-body Poincar\'e-covariant Faddeev equation approach to $N$, $\Delta$ elastic and transition form factors or numerical simulations of lQCD.

\medskip


\noindent\emph{Acknowledgments}\,---\,%
We are grateful for constructive communications with
K.~Graczyk, T.-S.\,H.~Lee, A.~Lovato and T.~Sato.
Work supported by:
National Natural Science Foundation of China (grant nos.\ 12135007, 12247103);
and
Deutsche Forschungsgemeinschaft (grant no.\ FI 970/11-1).


\begin{thebibliography}{58}
\providecommand{\natexlab}[1]{#1}
\providecommand{\url}[1]{\texttt{#1}}
\providecommand{\urlprefix}{URL }
\expandafter\ifx\csname urlstyle\endcsname\relax
  \providecommand{\doi}[1]{doi:\discretionary{}{}{}#1}\else
  \providecommand{\doi}[1]{doi:\discretionary{}{}{}\begingroup
  \urlstyle{rm}\url{#1}\endgroup}\fi
\providecommand{\bibinfo}[2]{#2}

\bibitem[{Fukuda et~al.(1998)}]{Super-Kamiokande:1998kpq}
\bibinfo{author}{Y.~Fukuda}, et~al., \bibinfo{title}{{Evidence for oscillation
  of atmospheric neutrinos}}, \bibinfo{journal}{Phys. Rev. Lett.}
  \bibinfo{volume}{81} (\bibinfo{year}{1998}) \bibinfo{pages}{1562--1567}.

\bibitem[{Kajita(2016)}]{RevModPhys.88.030501}
\bibinfo{author}{T.~Kajita}, \bibinfo{title}{Nobel Lecture: Discovery of
  atmospheric neutrino oscillations}, \bibinfo{journal}{Rev. Mod. Phys.}
  \bibinfo{volume}{88} (\bibinfo{year}{2016}) \bibinfo{pages}{030501}.

\bibitem[{McDonald(2016)}]{RevModPhys.88.030502}
\bibinfo{author}{A.~B. McDonald}, \bibinfo{title}{Nobel Lecture: The Sudbury
  Neutrino Observatory: Observation of flavor change for solar neutrinos},
  \bibinfo{journal}{Rev. Mod. Phys.} \bibinfo{volume}{88}
  (\bibinfo{year}{2016}) \bibinfo{pages}{030502}.

\bibitem[{Alvarez-Ruso et~al.(2018)}]{Alvarez-Ruso:2017oui}
\bibinfo{author}{L.~Alvarez-Ruso}, et~al., \bibinfo{title}{{NuSTEC White Paper:
  Status and challenges of neutrino\textendash{}nucleus scattering}},
  \bibinfo{journal}{Prog. Part. Nucl. Phys.} \bibinfo{volume}{100}
  (\bibinfo{year}{2018}) \bibinfo{pages}{1--68}.

\bibitem[{Arg\"uelles et~al.(2020)}]{Arguelles:2019xgp}
\bibinfo{author}{C.~A. Arg\"uelles}, et~al., \bibinfo{title}{{New opportunities
  at the next-generation neutrino experiments I: BSM neutrino physics and dark
  matter}}, \bibinfo{journal}{Rept. Prog. Phys.}
  \bibinfo{volume}{83}~(\bibinfo{number}{12}) (\bibinfo{year}{2020})
  \bibinfo{pages}{124201}.

\bibitem[{Abi et~al.(2020)}]{DUNE:2020jqi}
\bibinfo{author}{B.~Abi}, et~al., \bibinfo{title}{{Long-baseline neutrino
  oscillation physics potential of the DUNE experiment}},
  \bibinfo{journal}{Eur. Phys. J. C}
  \bibinfo{volume}{80}~(\bibinfo{number}{10}) (\bibinfo{year}{2020})
  \bibinfo{pages}{978}.

\bibitem[{Formaggio et~al.(2021)Formaggio, de~Gouv\^ea, and
  Robertson}]{Formaggio:2021nfz}
\bibinfo{author}{J.~A. Formaggio}, \bibinfo{author}{A.~L.~C. de~Gouv\^ea},
  \bibinfo{author}{R.~G.~H. Robertson}, \bibinfo{title}{{Direct Measurements of
  Neutrino Mass}}, \bibinfo{journal}{Phys. Rept.} \bibinfo{volume}{914}
  (\bibinfo{year}{2021}) \bibinfo{pages}{1--54}.

\bibitem[{Abusleme et~al.(2022)}]{JUNO:2021vlw}
\bibinfo{author}{A.~Abusleme}, et~al., \bibinfo{title}{{JUNO physics and
  detector}}, \bibinfo{journal}{Prog. Part. Nucl. Phys.} \bibinfo{volume}{123}
  (\bibinfo{year}{2022}) \bibinfo{pages}{103927}.

\bibitem[{Lokhov et~al.(2022)Lokhov, Mertens, Parno, Schl\"osser, and
  Valerius}]{Lokhov:2022zfn}
\bibinfo{author}{A.~Lokhov}, \bibinfo{author}{S.~Mertens},
  \bibinfo{author}{D.~S. Parno}, \bibinfo{author}{M.~Schl\"osser},
  \bibinfo{author}{K.~Valerius}, \bibinfo{title}{{Probing the Neutrino-Mass
  Scale with the KATRIN Experiment}}, \bibinfo{journal}{Ann. Rev. Nucl. Part.
  Sci.} \bibinfo{volume}{72} (\bibinfo{year}{2022}) \bibinfo{pages}{259--282}.

\bibitem[{Ruso et~al.(2022)}]{Ruso:2022qes}
\bibinfo{author}{L.~A. Ruso}, et~al., \bibinfo{title}{{Theoretical tools for
  neutrino scattering: interplay between lattice QCD, EFTs, nuclear physics,
  phenomenology, and neutrino event generators -- arXiv:2203.09030 [hep-ph]}} .

\bibitem[{Sajjad~Athar et~al.(2023)Sajjad~Athar, Fatima, and
  Singh}]{SajjadAthar:2022pjt}
\bibinfo{author}{M.~Sajjad~Athar}, \bibinfo{author}{A.~Fatima},
  \bibinfo{author}{S.~K. Singh}, \bibinfo{title}{{Neutrinos and their
  interactions with matter}}, \bibinfo{journal}{Prog. Part. Nucl. Phys.}
  \bibinfo{volume}{129} (\bibinfo{year}{2023}) \bibinfo{pages}{104019}.

\bibitem[{Mosel(2016)}]{Mosel:2016cwa}
\bibinfo{author}{U.~Mosel}, \bibinfo{title}{{Neutrino Interactions with
  Nucleons and Nuclei: Importance for Long-Baseline Experiments}},
  \bibinfo{journal}{Ann. Rev. Nucl. Part. Sci.} \bibinfo{volume}{66}
  (\bibinfo{year}{2016}) \bibinfo{pages}{171--195}.

\bibitem[{Hill et~al.(2018)Hill, Kammel, Marciano, and Sirlin}]{Hill:2017wgb}
\bibinfo{author}{R.~J. Hill}, \bibinfo{author}{P.~Kammel},
  \bibinfo{author}{W.~J. Marciano}, \bibinfo{author}{A.~Sirlin},
  \bibinfo{title}{{Nucleon Axial Radius and Muonic Hydrogen \textemdash{} A New
  Analysis and Review}}, \bibinfo{journal}{Rept. Prog. Phys.}
  \bibinfo{volume}{81} (\bibinfo{year}{2018}) \bibinfo{pages}{096301}.

\bibitem[{Gysbers et~al.(2019)}]{Gysbers:2019uyb}
\bibinfo{author}{P.~Gysbers}, et~al., \bibinfo{title}{{Discrepancy between
  experimental and theoretical $\beta$-decay rates resolved from first
  principles}}, \bibinfo{journal}{Nature Phys.}
  \bibinfo{volume}{15}~(\bibinfo{number}{5}) (\bibinfo{year}{2019})
  \bibinfo{pages}{428--431}.

\bibitem[{Lovato et~al.(2020)Lovato, Carlson, Gandolfi, Rocco, and
  Schiavilla}]{Lovato:2020kba}
\bibinfo{author}{A.~Lovato}, \bibinfo{author}{J.~Carlson},
  \bibinfo{author}{S.~Gandolfi}, \bibinfo{author}{N.~Rocco},
  \bibinfo{author}{R.~Schiavilla}, \bibinfo{title}{{Ab initio study of
  $\boldsymbol{(\nu_\ell,\ell^-)}$ and
  $\boldsymbol{(\overline{\nu}_\ell,\ell^+)}$ inclusive scattering in $^{12}$C:
  confronting the MiniBooNE and T2K CCQE data}}, \bibinfo{journal}{Phys. Rev.
  X} \bibinfo{volume}{10} (\bibinfo{year}{2020}) \bibinfo{pages}{031068}.

\bibitem[{Chen et~al.(2022)Chen, Fischer, Roberts, and Segovia}]{Chen:2021guo}
\bibinfo{author}{C.~Chen}, \bibinfo{author}{C.~S. Fischer},
  \bibinfo{author}{C.~D. Roberts}, \bibinfo{author}{J.~Segovia},
  \bibinfo{title}{{Nucleon axial-vector and pseudoscalar form factors and PCAC
  relations}}, \bibinfo{journal}{Phys. Rev. D}
  \bibinfo{volume}{105}~(\bibinfo{number}{9}) (\bibinfo{year}{2022})
  \bibinfo{pages}{094022}.

\bibitem[{Chen and Roberts(2022)}]{ChenChen:2022qpy}
\bibinfo{author}{C.~Chen}, \bibinfo{author}{C.~D. Roberts},
  \bibinfo{title}{{Nucleon axial form factor at large momentum transfers}},
  \bibinfo{journal}{Eur. Phys. J. A} \bibinfo{volume}{58}
  (\bibinfo{year}{2022}) \bibinfo{pages}{206}.

\bibitem[{Alexandrou et~al.(2017)Alexandrou, Constantinou, Hadjiyiannakou,
  Jansen, Kallidonis, Koutsou, and Vaquero Aviles-Casco}]{Alexandrou:2017hac}
\bibinfo{author}{C.~Alexandrou}, \bibinfo{author}{M.~Constantinou},
  \bibinfo{author}{K.~Hadjiyiannakou}, \bibinfo{author}{K.~Jansen},
  \bibinfo{author}{C.~Kallidonis}, \bibinfo{author}{G.~Koutsou},
  \bibinfo{author}{A.~Vaquero Aviles-Casco}, \bibinfo{title}{{Nucleon axial
  form factors using $N_f$ = 2 twisted mass fermions with a physical value of
  the pion mass}}, \bibinfo{journal}{Phys. Rev. D} \bibinfo{volume}{96}
  (\bibinfo{year}{2017}) \bibinfo{pages}{054507}.

\bibitem[{Jang et~al.(2020)Jang, Gupta, Yoon, and Bhattacharya}]{Jang:2019vkm}
\bibinfo{author}{Y.-C. Jang}, \bibinfo{author}{R.~Gupta},
  \bibinfo{author}{B.~Yoon}, \bibinfo{author}{T.~Bhattacharya},
  \bibinfo{title}{{Axial Vector Form Factors from Lattice QCD that Satisfy the
  PCAC Relation}}, \bibinfo{journal}{Phys. Rev. Lett.} \bibinfo{volume}{124}
  (\bibinfo{year}{2020}) \bibinfo{pages}{072002}.

\bibitem[{Bali et~al.(2020)Bali, Barca, Collins, Gruber, L{\"o}ffler,
  Sch{\"a}fer, S{\"o}ldner, Wein, Weish{\"a}upl, and Wurm}]{Bali:2019yiy}
\bibinfo{author}{G.~S. Bali}, \bibinfo{author}{L.~Barca},
  \bibinfo{author}{S.~Collins}, \bibinfo{author}{M.~Gruber},
  \bibinfo{author}{M.~L{\"o}ffler}, \bibinfo{author}{A.~Sch{\"a}fer},
  \bibinfo{author}{W.~S{\"o}ldner}, \bibinfo{author}{P.~Wein},
  \bibinfo{author}{S.~Weish{\"a}upl}, \bibinfo{author}{T.~Wurm},
  \bibinfo{title}{{Nucleon axial structure from lattice QCD}},
  \bibinfo{journal}{JHEP} \bibinfo{volume}{05} (\bibinfo{year}{2020})
  \bibinfo{pages}{126 (2020)}.

\bibitem[{Sato et~al.(2003)Sato, Uno, and Lee}]{Sato:2003rq}
\bibinfo{author}{T.~Sato}, \bibinfo{author}{D.~Uno}, \bibinfo{author}{T.~S.~H.
  Lee}, \bibinfo{title}{{Dynamical model of weak pion production reactions}},
  \bibinfo{journal}{Phys. Rev. C} \bibinfo{volume}{67} (\bibinfo{year}{2003})
  \bibinfo{pages}{065201}.

\bibitem[{Simons et~al.(2022)Simons, Steinberg, Lovato, Meurice, Rocco, and
  Wagman}]{Simons:2022ltq}
\bibinfo{author}{D.~Simons}, \bibinfo{author}{N.~Steinberg},
  \bibinfo{author}{A.~Lovato}, \bibinfo{author}{Y.~Meurice},
  \bibinfo{author}{N.~Rocco}, \bibinfo{author}{M.~Wagman},
  \bibinfo{title}{{Form factor and model dependence in neutrino-nucleus cross
  section predictions -- arXiv:2210.02455 [hep-ph]}} .

\bibitem[{Alexandrou et~al.(2007)Alexandrou, Koutsou, Leontiou, Negele, and
  Tsapalis}]{Alexandrou:2007eyf}
\bibinfo{author}{C.~Alexandrou}, \bibinfo{author}{G.~Koutsou},
  \bibinfo{author}{T.~Leontiou}, \bibinfo{author}{J.~W. Negele},
  \bibinfo{author}{A.~Tsapalis}, \bibinfo{title}{{Axial Nucleon and Nucleon to
  Delta form factors and the Goldberger-Treiman Relations from Lattice QCD}},
  \bibinfo{journal}{Phys. Rev. D} \bibinfo{volume}{76} (\bibinfo{year}{2007})
  \bibinfo{pages}{094511}, \bibinfo{note}{[Erratum: Phys. Rev. D 80, 099901
  (2009)]}.

\bibitem[{Alexandrou et~al.(2011)Alexandrou, Koutsou, Negele, Proestos, and
  Tsapalis}]{Alexandrou:2010uk}
\bibinfo{author}{C.~Alexandrou}, \bibinfo{author}{G.~Koutsou},
  \bibinfo{author}{J.~W. Negele}, \bibinfo{author}{Y.~Proestos},
  \bibinfo{author}{A.~Tsapalis}, \bibinfo{title}{{Nucleon to Delta transition
  form factors with $N_F=2+1$ domain wall fermions}}, \bibinfo{journal}{Phys.
  Rev. D} \bibinfo{volume}{83} (\bibinfo{year}{2011}) \bibinfo{pages}{014501}.

\bibitem[{Eichmann et~al.(2016)Eichmann, Sanchis-Alepuz, Williams, Alkofer, and
  Fischer}]{Eichmann:2016yit}
\bibinfo{author}{G.~Eichmann}, \bibinfo{author}{H.~Sanchis-Alepuz},
  \bibinfo{author}{R.~Williams}, \bibinfo{author}{R.~Alkofer},
  \bibinfo{author}{C.~S. Fischer}, \bibinfo{title}{{Baryons as relativistic
  three-quark bound states}}, \bibinfo{journal}{Prog. Part. Nucl. Phys.}
  \bibinfo{volume}{91} (\bibinfo{year}{2016}) \bibinfo{pages}{1--100}.

\bibitem[{Barabanov et~al.(2021)}]{Barabanov:2020jvn}
\bibinfo{author}{M.~Y. Barabanov}, et~al., \bibinfo{title}{{Diquark
  Correlations in Hadron Physics: Origin, Impact and Evidence}},
  \bibinfo{journal}{Prog. Part. Nucl. Phys.} \bibinfo{volume}{116}
  (\bibinfo{year}{2021}) \bibinfo{pages}{103835}.

\bibitem[{Ding et~al.(2023)Ding, Roberts, and Schmidt}]{Ding:2022ows}
\bibinfo{author}{M.~Ding}, \bibinfo{author}{C.~D. Roberts},
  \bibinfo{author}{S.~M. Schmidt}, \bibinfo{title}{{Emergence of Hadron Mass
  and Structure}}, \bibinfo{journal}{Particles}
  \bibinfo{volume}{6}~(\bibinfo{number}{1}) (\bibinfo{year}{2023})
  \bibinfo{pages}{57--120}.

\bibitem[{Carman et~al.(2023)Carman, Gothe, Mokeev, and
  Roberts}]{Carman:2023zke}
\bibinfo{author}{D.~S. Carman}, \bibinfo{author}{R.~W. Gothe},
  \bibinfo{author}{V.~I. Mokeev}, \bibinfo{author}{C.~D. Roberts},
  \bibinfo{title}{{Nucleon Resonance Electroexcitation Amplitudes and Emergent
  Hadron Mass}}, \bibinfo{journal}{Particles}
  \bibinfo{volume}{6}~(\bibinfo{number}{1}) (\bibinfo{year}{2023})
  \bibinfo{pages}{416--439}.

\bibitem[{Yin et~al.(2023)Yin, Chen, Fischer, and Roberts}]{Yin:2023kom}
\bibinfo{author}{P.-L. Yin}, \bibinfo{author}{C.~Chen}, \bibinfo{author}{C.~S.
  Fischer}, \bibinfo{author}{C.~D. Roberts}, \bibinfo{title}{{$\Delta $-Baryon
  axialvector and pseudoscalar form factors, and associated PCAC relations}},
  \bibinfo{journal}{Eur. Phys. J. A} \bibinfo{volume}{59}~(\bibinfo{number}{7})
  (\bibinfo{year}{2023}) \bibinfo{pages}{163}.

\bibitem[{Itzykson and Zuber(1980)}]{IZ80}
\bibinfo{author}{C.~Itzykson}, \bibinfo{author}{J.-B. Zuber},
  \bibinfo{title}{Quantum Field Theory}, \bibinfo{publisher}{McGraw-Hill Inc.,
  New York}, \bibinfo{year}{1980}.

\bibitem[{Segovia et~al.(2014)Segovia, Cloet, Roberts, and
  Schmidt}]{Segovia:2014aza}
\bibinfo{author}{J.~Segovia}, \bibinfo{author}{I.~C. Cloet},
  \bibinfo{author}{C.~D. Roberts}, \bibinfo{author}{S.~M. Schmidt},
  \bibinfo{title}{{Nucleon and $\Delta$ elastic and transition form factors}},
  \bibinfo{journal}{Few Body Syst.} \bibinfo{volume}{55} (\bibinfo{year}{2014})
  \bibinfo{pages}{1185--1222}.

\bibitem[{Adler(1968)}]{Adler:1968tw}
\bibinfo{author}{S.~L. Adler}, \bibinfo{title}{{Photoproduction,
  electroproduction and weak single pion production in the (3,3) resonance
  region}}, \bibinfo{journal}{Annals Phys.} \bibinfo{volume}{50}
  (\bibinfo{year}{1968}) \bibinfo{pages}{189--311}.

\bibitem[{Llewellyn~Smith(1972)}]{LlewellynSmith:1971uhs}
\bibinfo{author}{C.~H. Llewellyn~Smith}, \bibinfo{title}{{Neutrino Reactions at
  Accelerator Energies}}, \bibinfo{journal}{Phys. Rept.} \bibinfo{volume}{3}
  (\bibinfo{year}{1972}) \bibinfo{pages}{261--379}.

\bibitem[{Workman et~al.(2022)}]{Workman:2022ynf}
\bibinfo{author}{R.~L. Workman}, et~al., \bibinfo{title}{{Review of Particle
  Physics}}, \bibinfo{journal}{PTEP} \bibinfo{volume}{2022}
  (\bibinfo{year}{2022}) \bibinfo{pages}{083C01}.

\bibitem[{Cui et~al.(2020)Cui, Zhang, Binosi, de~Soto, Mezrag, Papavassiliou,
  Roberts, Rodr{\'{\i}}guez-Quintero, Segovia, and Zafeiropoulos}]{Cui:2019dwv}
\bibinfo{author}{Z.-F. Cui}, \bibinfo{author}{J.-L. Zhang},
  \bibinfo{author}{D.~Binosi}, \bibinfo{author}{F.~de~Soto},
  \bibinfo{author}{C.~Mezrag}, \bibinfo{author}{J.~Papavassiliou},
  \bibinfo{author}{C.~D. Roberts},
  \bibinfo{author}{J.~Rodr{\'{\i}}guez-Quintero}, \bibinfo{author}{J.~Segovia},
  \bibinfo{author}{S.~Zafeiropoulos}, \bibinfo{title}{{Effective charge from
  lattice QCD}}, \bibinfo{journal}{Chin. Phys. C} \bibinfo{volume}{44}
  (\bibinfo{year}{2020}) \bibinfo{pages}{083102}.

\bibitem[{Deur et~al.(2024)Deur, Brodsky, and Roberts}]{Deur:2023dzc}
\bibinfo{author}{A.~Deur}, \bibinfo{author}{S.~J. Brodsky},
  \bibinfo{author}{C.~D. Roberts}, \bibinfo{title}{{QCD Running Couplings and
  Effective Charges}}, \bibinfo{journal}{Prog. Part. Nucl. Phys.}
  \bibinfo{volume}{134} (\bibinfo{year}{2024}) \bibinfo{pages}{104081}.

\bibitem[{Cahill et~al.(1989)Cahill, Roberts, and Praschifka}]{Cahill:1988dx}
\bibinfo{author}{R.~T. Cahill}, \bibinfo{author}{C.~D. Roberts},
  \bibinfo{author}{J.~Praschifka}, \bibinfo{title}{{Baryon structure and QCD}},
  \bibinfo{journal}{Austral. J. Phys.} \bibinfo{volume}{42}
  (\bibinfo{year}{1989}) \bibinfo{pages}{129--145}.

\bibitem[{Reinhardt(1990)}]{Reinhardt:1989rw}
\bibinfo{author}{H.~Reinhardt}, \bibinfo{title}{{Hadronization of Quark Flavor
  Dynamics}}, \bibinfo{journal}{Phys. Lett. B} \bibinfo{volume}{244}
  (\bibinfo{year}{1990}) \bibinfo{pages}{316--326}.

\bibitem[{Efimov et~al.(1990)Efimov, Ivanov, and Lyubovitskij}]{Efimov:1990uz}
\bibinfo{author}{G.~V. Efimov}, \bibinfo{author}{M.~A. Ivanov},
  \bibinfo{author}{V.~E. Lyubovitskij}, \bibinfo{title}{{Quark - diquark
  approximation of the three quark structure of baryons in the quark
  confinement model}}, \bibinfo{journal}{Z. Phys. C} \bibinfo{volume}{47}
  (\bibinfo{year}{1990}) \bibinfo{pages}{583--594}.

\bibitem[{Roberts et~al.(2021)Roberts, Richards, Horn, and
  Chang}]{Roberts:2021nhw}
\bibinfo{author}{C.~D. Roberts}, \bibinfo{author}{D.~G. Richards},
  \bibinfo{author}{T.~Horn}, \bibinfo{author}{L.~Chang},
  \bibinfo{title}{{Insights into the emergence of mass from studies of pion and
  kaon structure}}, \bibinfo{journal}{Prog. Part. Nucl. Phys.}
  \bibinfo{volume}{120} (\bibinfo{year}{2021}) \bibinfo{pages}{103883}.

\bibitem[{Binosi(2022)}]{Binosi:2022djx}
\bibinfo{author}{D.~Binosi}, \bibinfo{title}{{Emergent Hadron Mass in Strong
  Dynamics}}, \bibinfo{journal}{Few Body Syst.}
  \bibinfo{volume}{63}~(\bibinfo{number}{2}) (\bibinfo{year}{2022})
  \bibinfo{pages}{42}.

\bibitem[{Salm\`e(2022)}]{Salme:2022eoy}
\bibinfo{author}{G.~Salm\`e}, \bibinfo{title}{{Explaining mass and spin in the
  visible matter: the next challenge}}, \bibinfo{journal}{J. Phys. Conf. Ser.}
  \bibinfo{volume}{2340}~(\bibinfo{number}{1}) (\bibinfo{year}{2022})
  \bibinfo{pages}{012011}.

\bibitem[{de~Teramond(2022)}]{deTeramond:2022zcm}
\bibinfo{author}{G.~F. de~Teramond}, \bibinfo{title}{{Emergent phenomena in
  QCD: The holographic perspective -- arXiv:2212.14028 [hep-ph]}}, in:
  \bibinfo{booktitle}{{25th Workshop on What Comes Beyond the Standard
  Models?}}, \bibinfo{year}{2022}.

\bibitem[{Ferreira and Papavassiliou(2023)}]{Ferreira:2023fva}
\bibinfo{author}{M.~N. Ferreira}, \bibinfo{author}{J.~Papavassiliou},
  \bibinfo{title}{{Gauge Sector Dynamics in QCD}}, \bibinfo{journal}{Particles}
  \bibinfo{volume}{6}~(\bibinfo{number}{1}) (\bibinfo{year}{2023})
  \bibinfo{pages}{312--363}.

\bibitem[{Krein(2023)}]{Krein:2023azg}
\bibinfo{author}{G.~Krein}, \bibinfo{title}{{Femtoscopy of the Matter
  Distribution in the Proton}}, \bibinfo{journal}{Few Body Syst.}
  \bibinfo{volume}{64}~(\bibinfo{number}{3}) (\bibinfo{year}{2023})
  \bibinfo{pages}{42}.

\bibitem[{Kitagaki et~al.(1990)}]{Kitagaki:1990vs}
\bibinfo{author}{T.~Kitagaki}, et~al., \bibinfo{title}{{Study of $\nu d \to
  \mu^- pp_s$ and $\nu d \to \mu^- \Delta^{++} n_s$ using the BNL 7-foot
  deuterium filled bubble chamber}}, \bibinfo{journal}{Phys. Rev. D}
  \bibinfo{volume}{42} (\bibinfo{year}{1990}) \bibinfo{pages}{1331--1338}.

\bibitem[{Schreiner and Von~Hippel(1973)}]{Schreiner:1973mj}
\bibinfo{author}{P.~A. Schreiner}, \bibinfo{author}{F.~Von~Hippel},
  \bibinfo{title}{{Neutrino production of the Delta (1236)}},
  \bibinfo{journal}{Nucl. Phys. B} \bibinfo{volume}{58} (\bibinfo{year}{1973})
  \bibinfo{pages}{333--362}.

\bibitem[{Meyer et~al.(2016)Meyer, Betancourt, Gran, and Hill}]{Meyer:2016oeg}
\bibinfo{author}{A.~S. Meyer}, \bibinfo{author}{M.~Betancourt},
  \bibinfo{author}{R.~Gran}, \bibinfo{author}{R.~J. Hill},
  \bibinfo{title}{{Deuterium target data for precision neutrino-nucleus cross
  sections}}, \bibinfo{journal}{Phys. Rev. D} \bibinfo{volume}{93}
  (\bibinfo{year}{2016}) \bibinfo{pages}{113015}.

\bibitem[{Liu et~al.(1995)Liu, Mukhopadhyay, and Zhang}]{Liu:1995bu}
\bibinfo{author}{J.~Liu}, \bibinfo{author}{N.~C. Mukhopadhyay},
  \bibinfo{author}{L.-s. Zhang}, \bibinfo{title}{{Nucleon to delta weak
  excitation amplitudes in the nonrelativistic quark model}},
  \bibinfo{journal}{Phys. Rev. C} \bibinfo{volume}{52} (\bibinfo{year}{1995})
  \bibinfo{pages}{1630--1647}.

\bibitem[{Barquilla-Cano et~al.(2007)Barquilla-Cano, Buchmann, and
  Hernandez}]{Barquilla-Cano:2007vds}
\bibinfo{author}{D.~Barquilla-Cano}, \bibinfo{author}{A.~J. Buchmann},
  \bibinfo{author}{E.~Hernandez}, \bibinfo{title}{{Axial $N\to \Delta(1232)$
  and $N\to N^\ast(1440)$ transition form factors}}, \bibinfo{journal}{Phys.
  Rev. C} \bibinfo{volume}{75} (\bibinfo{year}{2007}) \bibinfo{pages}{065203},
  \bibinfo{note}{[Erratum: Phys.Rev.C 77, 019903 (2008)]}.

\bibitem[{Liu et~al.(2022)Liu, Chen, Lu, Roberts, and Segovia}]{Liu:2022ndb}
\bibinfo{author}{L.~Liu}, \bibinfo{author}{C.~Chen}, \bibinfo{author}{Y.~Lu},
  \bibinfo{author}{C.~D. Roberts}, \bibinfo{author}{J.~Segovia},
  \bibinfo{title}{{Composition of low-lying $J=\tfrac{3}{2}^\pm$
  \ensuremath{\Delta}-baryons}}, \bibinfo{journal}{Phys. Rev. D}
  \bibinfo{volume}{105}~(\bibinfo{number}{11}) (\bibinfo{year}{2022})
  \bibinfo{pages}{114047}.

\bibitem[{Androic et~al.(2011)}]{G0:2011rpu}
\bibinfo{author}{D.~Androic}, et~al., \bibinfo{title}{{The G0 Experiment:
  Apparatus for Parity-Violating Electron Scattering Measurements at Forward
  and Backward Angles}}, \bibinfo{journal}{Nucl. Instrum. Meth. A}
  \bibinfo{volume}{646} (\bibinfo{year}{2011}) \bibinfo{pages}{59--86}.

\bibitem[{Becker et~al.(2018)}]{Becker:2018ggl}
\bibinfo{author}{D.~Becker}, et~al., \bibinfo{title}{{The P2 experiment}},
  \bibinfo{journal}{Eur. Phys. J. A}
  \bibinfo{volume}{54}~(\bibinfo{number}{11}) (\bibinfo{year}{2018})
  \bibinfo{pages}{208}.

\bibitem[{Arrington et~al.(2023)}]{JeffersonLabSoLID:2022iod}
\bibinfo{author}{J.~Arrington}, et~al., \bibinfo{title}{{The solenoidal large
  intensity device (SoLID) for JLab 12 GeV}}, \bibinfo{journal}{J. Phys. G}
  \bibinfo{volume}{50}~(\bibinfo{number}{11}) (\bibinfo{year}{2023})
  \bibinfo{pages}{110501}.

\bibitem[{Mukhopadhyay et~al.(1998)Mukhopadhyay, Ramsey-Musolf, Pollock, Liu,
  and Hammer}]{Mukhopadhyay:1998mn}
\bibinfo{author}{N.~C. Mukhopadhyay}, \bibinfo{author}{M.~J. Ramsey-Musolf},
  \bibinfo{author}{S.~J. Pollock}, \bibinfo{author}{J.~Liu},
  \bibinfo{author}{H.~W. Hammer}, \bibinfo{title}{{Parity violating excitation
  of the Delta (1232): Hadron structure and new physics}},
  \bibinfo{journal}{Nucl. Phys. A} \bibinfo{volume}{633} (\bibinfo{year}{1998})
  \bibinfo{pages}{481--518}.

\bibitem[{Julia-Diaz et~al.(2007)Julia-Diaz, Lee, Sato, and
  Smith}]{Julia-Diaz:2006ios}
\bibinfo{author}{B.~Julia-Diaz}, \bibinfo{author}{T.~S.~H. Lee},
  \bibinfo{author}{T.~Sato}, \bibinfo{author}{L.~C. Smith},
  \bibinfo{title}{{Extraction and Interpretation of $\gamma N \to \Delta$ Form
  Factors within a Dynamical Model}}, \bibinfo{journal}{Phys. Rev. C}
  \bibinfo{volume}{75} (\bibinfo{year}{2007}) \bibinfo{pages}{015205}.

\bibitem[{Drechsel et~al.(2007)Drechsel, Kamalov, and Tiator}]{Drechsel:2007if}
\bibinfo{author}{D.~Drechsel}, \bibinfo{author}{S.~S. Kamalov},
  \bibinfo{author}{L.~Tiator}, \bibinfo{title}{{Unitary Isobar Model -
  MAID2007}}, \bibinfo{journal}{Eur. Phys. J. A} \bibinfo{volume}{34}
  (\bibinfo{year}{2007}) \bibinfo{pages}{69--97}.

\bibitem[{\"Unal et~al.(2021)\"Unal, K\"u\c{c}\"ukarslan, and
  Scherer}]{Unal:2021byi}
\bibinfo{author}{Y.~\"Unal}, \bibinfo{author}{A.~K\"u\c{c}\"ukarslan},
  \bibinfo{author}{S.~Scherer}, \bibinfo{title}{{Axial-vector nucleon-to-delta
  transition form factors using the complex-mass renormalization scheme}},
  \bibinfo{journal}{Phys. Rev. D} \bibinfo{volume}{104}~(\bibinfo{number}{9})
  (\bibinfo{year}{2021}) \bibinfo{pages}{094014}.

\end{thebibliography}

\medskip

\newpage

\hspace*{-\parindent}\dotfill\ \emph{Supplemental Material} \dotfill

\smallskip

%
This material is included with a view to illustrating/exemplifying remarks in the main text, including, in some cases, numerical confirmation of observations inferred from figures, and supplying practicable representations of the form factor predictions.

\begin{figure}[t]
\centerline{\includegraphics[clip, height=0.33\textwidth, width=0.45\textwidth]{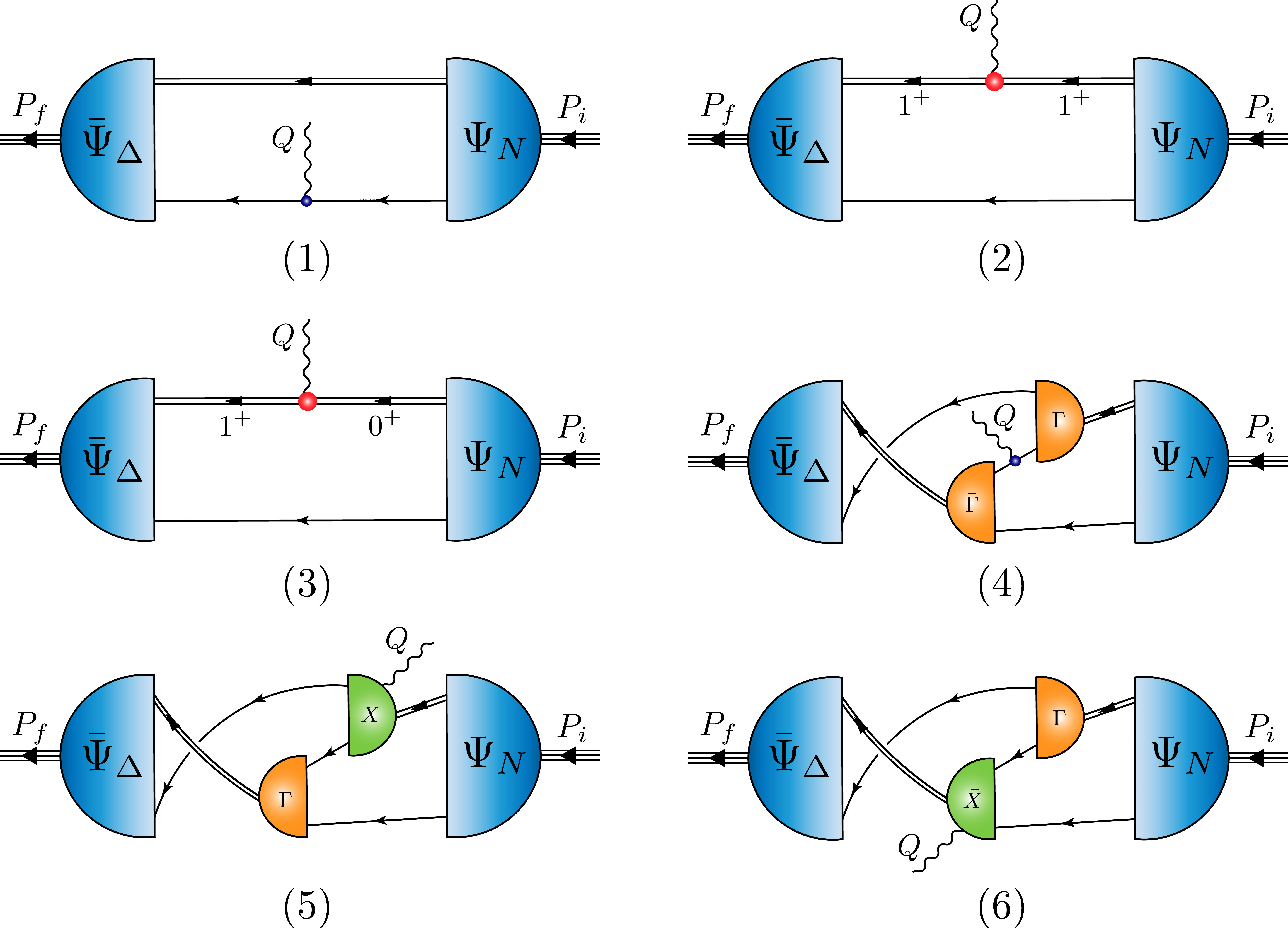}}
\caption{\label{figcurrent} Symmetry-preserving axial/pseudoscalar current for on-shell baryons ($B=N,\Delta$) that are described by the Faddeev amplitudes $\Psi_B$ obtained by solving the equation depicted in Fig.\,\ref{figFaddeev}. \emph{Single line} -- dressed-quark propagator; \emph{undulating line} -- axial/pseudoscalar current; $\Gamma$ -- diquark correlation amplitude; \emph{double line} -- diquark propagator; and $\chi$ -- seagull terms. }
\end{figure}

Within the quark + diquark picture of baryons, the axial and pseudoscalar currents, Eqs.\,\eqref{jaxdq0} and \eqref{jpsdq0}, can both be decomposed into a sum of six terms, depicted in Fig.~\ref{figcurrent}.  Evidently, the probe interacts with the dressed-quarks and -diquarks in various ways, each of which is detailed in Ref.\,\cite[Sec.\,IIIC]{Chen:2021guo}.
Referring to Fig.\,\ref{figcurrent}, in Table\,\ref{tabQ20sep} we list the relative strength of each diagram's contribution to $C^A_{3,4,5,6}(0)$ and $G_{\pi N \Delta}(0)$.
%

\begin{table}[t]
\caption{\label{tabQ20sep}
Referring to Fig.\,\ref{figcurrent}, separation of the axial-vector form factors $C^A_i(0)$ $(i=3,4,5,6)$, and the $\pi N\Delta$ form factor $G_{\pi N\Delta}(0)$ into contributions from various diagrams, listed as a fraction of the total $Q^2=0$ value.
Diagram (1): $\langle J \rangle^{A}_{\rm q}$ -- weak- or pseudoscalar-probe strikes dressed-quark with axialvector diquark spectator.
Diagram (2): $\langle J \rangle^{AA}_{\rm qq}$ -- probe strikes axialvector diquark with dressed-quark spectator.
Diagram (3): $\langle J \rangle^{AS}_{\rm qq}$ -- probe mediates scalar to axialvector diquark transition, with dressed-quark spectator.
Diagram (4): $\langle J \rangle_{\rm ex}$ -- probe strikes dressed-quark ``in-flight'' between one diquark correlation and another.
Diagrams (5) and (6): $\langle J \rangle_{\rm sg}$ -- probe couples inside the diquark correlation amplitude.
The listed uncertainties reflect the impact of $\pm 5$\% variations in the diquark masses in Eq.\,\eqref{dqmass}, \emph{e.g}., $-0.11_{6_\mp} \Rightarrow -0.11 \mp 0.06$.
}
\begin{center}
\begin{tabular*}
{\hsize}
{
l@{\extracolsep{0ptplus1fil}}
|l@{\extracolsep{0ptplus1fil}}
l@{\extracolsep{0ptplus1fil}}
l@{\extracolsep{0ptplus1fil}}
l@{\extracolsep{0ptplus1fil}}
l@{\extracolsep{0ptplus1fil}}}\hline\hline
 & $\ \langle J \rangle^{A}_{\rm q}$ &$\langle J \rangle^{AA}_{\rm qq}$ &$\langle J \rangle^{AS}_{\rm qq}$ & $\langle J \rangle_{\rm ex}$ & $\ \langle J \rangle_{\rm sg}$ \\\hline
 $C^A_5(0)\ $ & $\phantom{-}0.48_{9_\pm}$ & $0.017_{5_\pm}$ & $0.21_{1_\pm}$ & $0.29_{10_\mp}$ & $\phantom{-}0$\\
 $C^A_6(0)\ $ & $\phantom{-}0.50_{9_\pm}$ & $0.016_{5_\pm}$ & $0.21_{1_\pm}$ & $0.52_{8_\mp}$ & $-0.25_{2_\mp}$\\
 $C^A_3(0)\ $ & $-0.11_{5_\pm}$ & $0.14_{12_\pm}$ & $0.37_{10_\pm}$ & $0.60_{27_\mp}$ & $\phantom{-}0$\\
 $C^A_4(0)\ $ & $\phantom{-}0.44_{8_\pm}$ & $0.052_{37_\pm}$ & $0.23_{2_\pm}$ & $0.28_{10_\mp}$ & $\phantom{-}0$\\\hline
 $G_{\pi N\Delta}(0)\ $ & $\phantom{-}0.53_{11_\pm}$ & $0.020_{4_\pm}$ & $0.17_{2_\pm}$ & $0.57_{10_\mp}$ & $-0.29_{3_\mp}$\\[0.5ex]  \hline\hline
\end{tabular*}
\end{center}
\end{table}

In order to assist others in employing our predictions for the form factors associated with nucleon--to--$\Delta$ axial and pseudoscalar transition currents, here we present practicable parametrisations, valid as interpolations and useful for extrapolation.  We considered a range of options, achieving good results with the following functional forms:
\begin{equation}
\label{axpadeN}
{\mathscr F}(x) =\frac{n^0_{\mathscr F} + n^1_{\mathscr F} x / (n^2_{\mathscr F} +x)^2 }{(1 + d^1_{\mathscr F} x)^2}\,,
\end{equation}
for ${\mathscr F} =C^A_{i=3,4,5}$;
\begin{equation}
\label{pspade}
C^A_6(x)  =
\frac{n^0_{C^A_6} + n^1_{C^A_6} x / (n^2_{C^A_6} +x)^2 }{(1 + d^1_{C^A_6} x)^2}
 \frac{\mu_N}{x+\mu_\pi}\,,
\end{equation}
reflecting PPD, Eq.\,\eqref{ppd}; and
\begin{equation}
\label{axpade}
G_{\pi N\Delta}/m_N =
\frac{n^0_{G} + n^1_{G} x }{1 + d^1_{G} x + d^2_{G} x^2 + d^3_{G} x^3}\,.
\end{equation}
All interpolation coefficients are listed in Table\,\ref{tablepade}.

\begin{table}[t]
\caption{\label{tablepade}
Interpolation parameters for the $N\to\Delta$ axial and pseudoscalar transition form factors, for use in Eqs.\,\eqref{axpade}, \eqref{pspade}.  All dimensionless, except $n^{0,1}_G$, the numerator coefficients for  $G_{\pi\Delta\Delta}/m_N$ in Eq.\,\eqref{axpade}, which are listed in GeV$^{-1}$.
}
\begin{tabular*}
{\hsize}
{
l@{\extracolsep{0ptplus1fil}}
|c@{\extracolsep{0ptplus1fil}}
c@{\extracolsep{0ptplus1fil}}
c@{\extracolsep{0ptplus1fil}}
c@{\extracolsep{0ptplus1fil}}
c@{\extracolsep{0ptplus1fil}}}\hline\hline
 &  $C^A_5$ &  $C^A_6$ &  $C^A_3$ &  $C^A_4$ & $G_{\pi N\Delta}/m_N$\\\hline
 $n^0\ $ & $\phantom{-}1.16^{+0.09}_{-0.03}$ & $\phantom{-}1.43^{+0.17}_{-0.10}$ & $\phantom{-}0.26^{+0.17}_{-0.04}$ & $-0.66^{+0.03}_{-0.10}$
 & $\phantom{-}25.15^{+1.90}_{-0.67}$\\
 $n^1\ $ & $-0.27^{-0.19}_{+0.07}$ & $-0.22^{-0.13}_{+0.03}$ & $-0.87^{+0.34}_{+0.35}$ & $\phantom{-}7.85^{-1.22}_{-0.84}$
 & $-30.29^{+0.34}_{-1.27}$\\
 $n^2\ $ & $\phantom{-}0.46^{+0.05}_{-0.07}$ & $\phantom{-}0.17^{+0.02}_{+0.03}$ & $\phantom{-}2.02^{-1.25}_{+0.68}$ & $\phantom{-}3.67^{-0.35}_{-0.46}$ & \\
 $d^1\ $ & $\phantom{-}0.87^{-0.06}_{+0.06}$ & $\phantom{-}1.06^{-0.01}_{+0.01}$ & $\phantom{-}0.62^{+0.33}_{+0.17}$ & $\phantom{-}0.70^{-0.12}_{+0.24}$
 & $\phantom{-}3.31^{+0.01}_{+0.15}$\\
 $d^2\ $ & & & &
 & $\phantom{-}2.53^{+0.22}_{+0.15}$\\
 $d^3\ $ & & & &
 & $\phantom{-}1.88^{-0.58}_{+0.36}$\\\hline\hline
\end{tabular*}
\end{table}

\begin{figure}[t]
\centerline{%
\includegraphics[clip, width=0.45\textwidth]{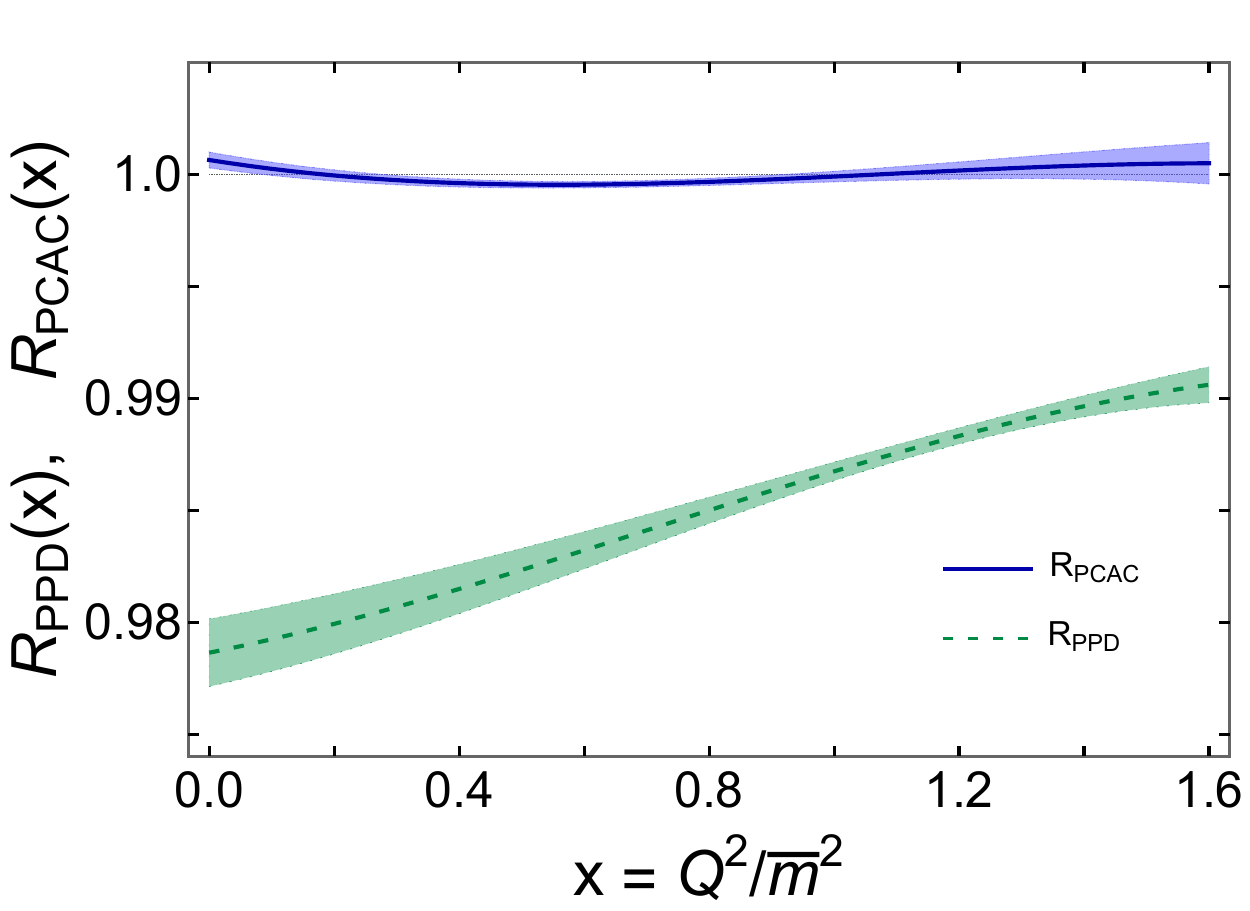}}
\caption{\label{figrpcac}
Quantitative verification of the accuracy of the PPD approximation -- Eq.\,\eqref{rppdnum}: dashed green curve within lighter band;
and
numerical verification of the PCAC relation -- Eq.\,\eqref{rpcacnum}: solid blue curve within lighter band.
}
\end{figure}

The off-diagonal PPD approximation is written in Eq.\,\eqref{ppd}.  Here we choose to supplement Fig.\,\ref{FigCA56x} with additional quantitative information.  Consider, therefore, the following ratio
\begin{align}
\label{rppdnum}
R_{PPD} := \frac{\mu_N C^A_5}{(x+\mu_\pi)C^A_6}\,.
\end{align}
The calculated result is depicted in Fig.\,\ref{figrpcac}.
The $x$-dependence of $R_{PPD}$ is very much like the analogous curves obtained for the nucleon \cite[Fig.\,8]{Chen:2021guo} and the $\Delta(1232)$-baryon \cite[Fig.\,8]{Yin:2023kom}: it lies $< 2.5\%$ below unity on $x\simeq 0$ and grows toward unity as $x$ increases.  The explanation for the accuracy and behaviour of this approximating formula can be found, \emph{e.g}., in Ref.\,\cite[Sec.\,4.3]{Yin:2023kom}.

Some corollaries of PCAC were discussed in connection with Eqs.\,\eqref{GTtest}, \eqref{psgt}.  As done separately for the nucleon and $\Delta$-baryon weak elastic form factors \cite{Chen:2021guo, Yin:2023kom}, they may be extended.  Indeed, following Ref.\,\cite[Sec.\,4.3]{Yin:2023kom}, one can prove algebraically that $C^A_5$, $C^A_6$, and $G_{\pi N\Delta}$ satisfy the off-diagonal PCAC relation, Eq.\,\eqref{pcacff}, as they must in any symmetry preserving treatment of $N\to \Delta$ weak transitions.  Thus, the PCAC ratio
\begin{align}
\label{rpcacnum}
R_{PCAC}:=\frac{2 m_N \mu_N C^A_5}{2m_N x C^A_6 + \mu_N\mu_\pi f_\pi G_{\pi N\Delta}/(x+\mu_\pi)}\,,
\end{align}
should be unity.  We draw the numerical result for this ratio in Fig.\,\ref{figrpcac}: evidently, allowing for numerical accuracy, the curve is indistinguishable from unity.  The uncertainty band highlights that the result is parameter-independent.

\end{document}